\documentclass[12pt,a4paper]{article}

\usepackage{epsfig}
\usepackage{amsmath,amsfonts,amssymb}
\usepackage{cite}
\usepackage{colortbl}
\usepackage{pifont}

\providecommand{\openone}{\leavevmode\hbox{\small1\kern-3.8pt\normalsize1}}

\begin{document}

\begin{center}
\begin{Large}
{\bf Profile of multiboson signals}
\end{Large}

\vspace{0.5cm}
J.~A.~Aguilar--Saavedra \\
\begin{small}
{ Departamento de F\'{\i}sica Te\'orica y del Cosmos, 
Universidad de Granada, \\ E-18071 Granada, Spain} \\ 

\end{small}
\end{center}

\begin{abstract}
We investigate the visibility of signals characterised by wide bumps over a smoothly falling background that cannot be accurately predicted by Monte Carlo calculations. Examples of such are the wide bumps that triboson and quadriboson resonance cascade decays would yield in diboson resonance searches in fully hadronic final states. We find that the sensitivity to triboson bumps is rather small: signals of a moderate size could be present in current data and yet remain unnoticed. For quadriboson cascade decays the signals can hardly be distinguished from the background in the current searches.
\end{abstract}

\section{Introduction}

\begin{figure}[htb]
\begin{center}
\includegraphics[height=3cm,clip=]{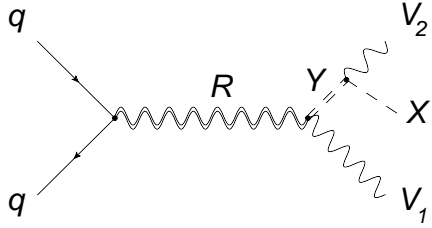}
\caption{Sample diagram for a $R \to V_1 Y \to V_1 V_2 X$ triboson signal.}
\label{fig:diag}
\end{center}
\end{figure}

In the absence of convincing signals of new physics at the Large Hadron Collider (LHC), the attention needs to be turned to non-minimal models giving non-standard signatures that are easy to miss by conventional searches. One example of such unexplored signals is a triboson resonance, that is, a cascade decay of a heavy resonance $R$ into a (massive) vector boson $V_1$ and an intermediate resonance $Y$, which subsequently decays into another vector boson $V_2$ plus an extra particle $X$. The latter particle can also be a vector boson, a Higgs boson, or a new particle, as depicted in figure~\ref{fig:diag}. This type of signal was introduced in ref.~\cite{Aguilar-Saavedra:2015rna} to explain a $3.4\sigma$ excess in a search of hadronically decaying diboson resonances by the ATLAS Collaboration in Run 1 data~\cite{Aad:2015owa} that was much milder in the analogous CMS search~\cite{Khachatryan:2014hpa} and did not show up in semileptonic final states~\cite{Aad:2015ufa,Khachatryan:2014gha}. Despite the ATLAS and CMS searches in hadronic final states had similar performance for true diboson resonances, this was not the case for a non-diboson signal, e.g. a triboson: the ATLAS event selection criteria would make it show up as a peak in the $V_1 V_2$ diboson invariant mass distribution while for the CMS analysis it would appear as a wide bump, harder to notice over the background. In addition, the presence of the extra particle $X$ could drastically reduce the signal efficiency for the searches in semileptonic final states.

\begin{figure}[htb]
\begin{center}
\includegraphics[height=3cm,clip=]{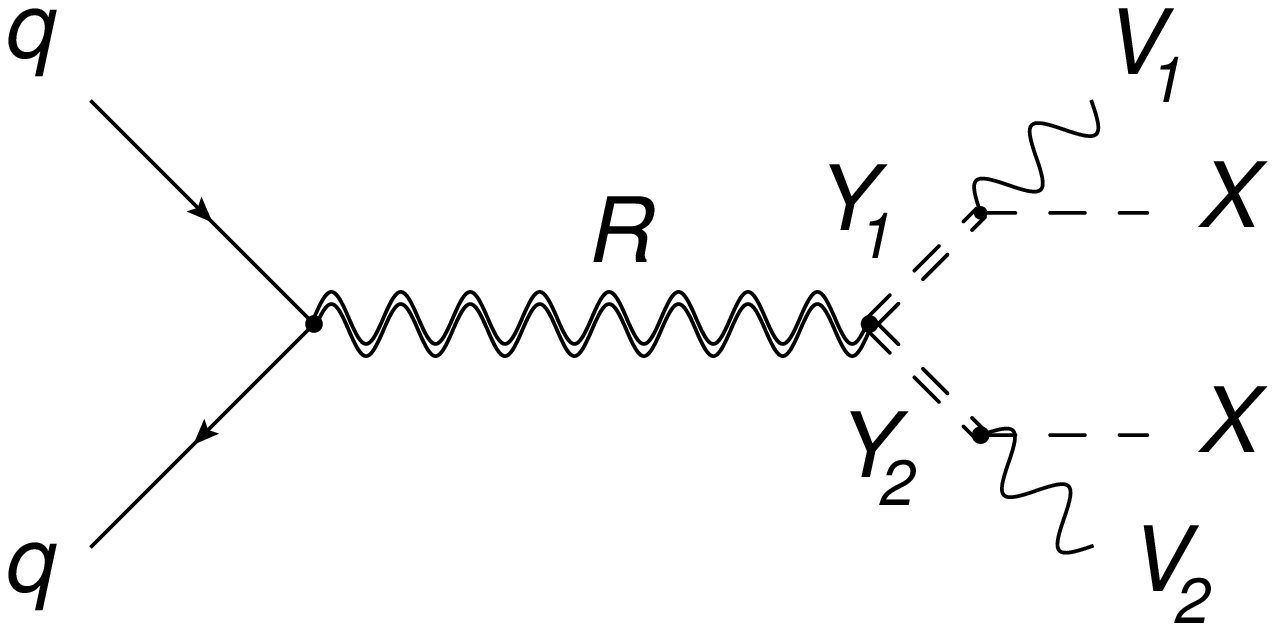}
\caption{Sample diagram for a $R \to Y_1 Y_2 \to V_1 V_2 XX$ quadriboson signal.}
\label{fig:diag2}
\end{center}
\end{figure}

Subsequently, it was shown that triboson signals can arise in the context of left-right models~\cite{Aguilar-Saavedra:2015iew}, with the new charged boson $W'$ being the heavy resonance $R$, and the extra scalars $H_1^0$, $H^\pm$ playing the role of the intermediate resonance $Y$. The extra particle $X$ can be a vector boson, the Higgs boson or the pseudo-scalar $A^0$.  For example, one can have $W' \to H^\pm Z \to W Z A^0$, or $W' \to H_1^0 W \to W Z A^0$. Quadriboson signals are also possible, with the heavy resonance ($W'$ or also $Z'$ in this context) decaying into two heavy scalars, which subsequently decay as depicted in figure~\ref{fig:diag2}. For example, one can have $W' \to H^\pm H_1^0 \to W Z A^0 A^0$.  

A more detailed simulation~\cite{Aguilar-Saavedra:2016xuc} confirmed that a potential $WZX$ triboson signal would be hard to see in semileptonic final states due to the low signal efficiency that results because the analyses are highly optimised for the kinematics of diboson resonances produced back-to-back in the transverse plane. Moreover, the event selection criteria often veto the presence of extra particles near the decay products of the boson with leptonic decay, to suppress SM backgrounds such as $t \bar t$ and $W/Z$ plus jets. This obviously removes a large fraction of a $V_1 V_2 X$ signal. Also, in the fully leptonic channels the branching ratios are quite small, and the produced signals are tiny. (Note that if the extra particle $X$ is not a SM gauge boson and decays hadronically, the only leptons result from the decay of the $W$ and $Z$ bosons.) The analysis of ref.~\cite{Aguilar-Saavedra:2016xuc} also confirmed that
in hadronic final states a significant peak shaping would be produced by the ATLAS event selection~\cite{Aad:2015owa,Aaboud:2016okv} but not in the CMS analyses~\cite{Khachatryan:2014hpa,Sirunyan:2016cao}, where the excess in the diboson invariant mass distribution would remain as a wide bump.  As it has been argued before~\cite{Aguilar-Saavedra:2015rna}, such wide bumps are not easy to spot in searches where the background itself is determined from data. 
In this work we address this specific point in detail, whose interest extends far beyond the interpretation of possible anomalies. In fact, as pointed out in ref.~\cite{Aguilar-Saavedra:2015iew}, triboson and quadriboson resonances are a natural possibility in models with several extra particles at different scales, with left-right models being only an example. And, as we will see, wide excesses such as those caused by multiboson resonances are easily absorbed in the (unknown) normalisation of the background. We begin by describing in section~\ref{sec:2} the typical profiles of the diboson invariant mass distributions for triboson and quadriboson signals. After this, in section~\ref{sec:3} we discuss the framework used to set limits on possible new resonances. In section~\ref{sec:4} we show to which extent the standard diboson searches are (in)sensitive to wide bumps caused by multiboson signals, parameterised by typical shapes obtained from simulation, and we compare to what one would expect for resonances with a Gaussian shape. We summarise our results and discuss their implications in section~\ref{sec:5}.


\section{Triboson and quadriboson shapes}
\label{sec:2}

\begin{figure}[htb]
\begin{center}
\includegraphics[height=5cm,clip=]{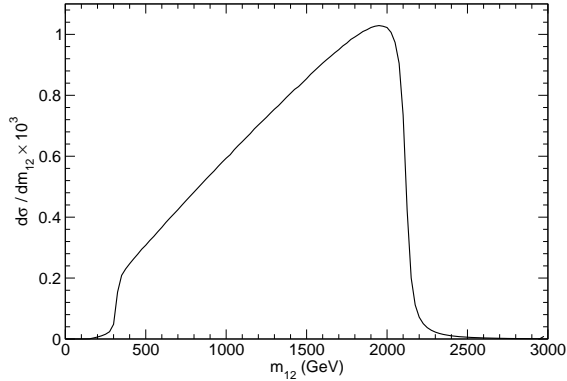}
\caption{Sample diboson invariant mass distribution $m_{12}$  for a $R \to V_1 Y \to V_1 V_2 X$ triboson signal at the partonic level, normalised to unity.}
\label{fig:dist-tribp}
\end{center}
\end{figure}

For triboson resonance cascade decays, such as the one depicted in figure~\ref{fig:diag},
the invariant mass distribution $m_{12}$ of the decay products $V_1$ and $V_2$ is very wide at the partonic level. An example is shown in figure~\ref{fig:dist-tribp}, taken for resonance masses $M_R = 2.25$ TeV, $M_Y = 650$ GeV, and boson mases $m_1 = M_W$, $m_2 = M_Z$, $m_X = 100$ GeV.\footnote{The value $M_Y = 650$ GeV assumed in ref.~\cite{Aguilar-Saavedra:2016xuc} was inspired by an excess in a search for $ZV$ resonances at this mass~\cite{CMS:2016tio}, which has not been confirmed with more data~\cite{CMS:2017mrw,CMS:2017sbi}. We take this value merely for illustration and consistency with earlier work.} Neglecting width effects and working at first order in $m_1^2$, $m_2^2$ and $m_X^2$, the minimum and maximum values are
\begin{align}
& (m_{12}^2)^\text{min} = M_R^2 \left( \frac{m_X^2}{M_Y^2} + \frac{m_2^2}{M_R^2 - M_Y^2} \right) + m_X^2 - m_1^2 \,, \notag \\
& (m_{12}^2)^\text{max} = M_R^2 \left( 1 - \frac{m_1^2}{M_Y^2} \right) - M_Y^2 \left(1+ \frac{m_2^2}{M_R^2 - M_Y^2} \right) + 2 m_X^2 \,.
\end{align}
If the intermediate particle is allowed to be very wide these values are shifted, recovering the upper limit for a three-body decay in ref.~\cite{Aguilar-Saavedra:2015rna}.

\begin{figure}[htb]
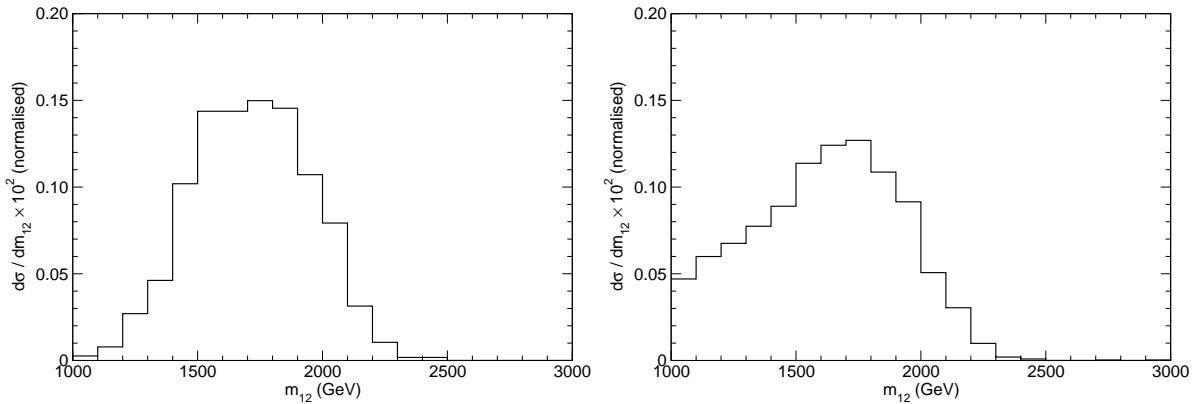

\begin{center}
\begin{tabular}{cc}
\includegraphics[height=5.25cm,clip=]{Figs/dist-bump1.eps}
\includegraphics[height=5.25cm,clip=]{Figs/dist-bump2.eps} \\
\end{tabular}
\caption{Sample distributions for triboson signals after event selection (see the text), normalised to unity. Left: symmetric bump. Right: asymmetric bump.}
\label{fig:dist-trib}
\end{center}
\end{figure}

In actual diboson resonance searches the invariant mass distribution of a potential triboson signal is shaped in varying degrees by the selection criteria applied on the transverse momentum $p_T$, rapidity $y_{1,2}$, rapidity difference $|y_1 - y_2|$, and $p_T$ balance of the bosons, as well as by the jet quality requirements. As aforementioned, in this paper we focus on the situations where the resulting signal distribution is a wide bump, and so we have selected two representative distributions that were obtained in ref.~\cite{Aguilar-Saavedra:2016xuc} after the event selection and reconstruction criteria. Figure~\ref{fig:dist-trib} (left) shows a symmetric bump obtained with the event selection of the ATLAS $VH$ search in the hadronic final state~\cite{ATLAS:2016kxc}, namely for the $WH$ selection. The right panel shows an asymmetric bump obtained with the high-purity $WZ$ event selection of the CMS Run 2 diboson search (2015 data) in the hadronic final state~\cite{Sirunyan:2016cao}. These two triboson signal shapes will be used in our subsequent analyses.

\begin{figure}[htb]
\begin{center}
\includegraphics[height=5.25cm,clip=]{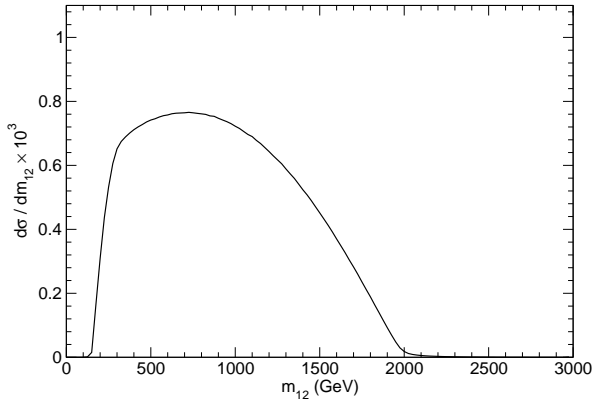}
\caption{Sample diboson invariant mass distribution $m_{12}$  for a $R \to Y_1 Y_2 \to V_1 V_2 X X$ quadriboson signal at the partonic level, normalised to unity.}
\label{fig:dist-quadp}
\end{center}
\end{figure}

\begin{figure}[htb]
\begin{center}
\includegraphics[height=5.25cm,clip=]{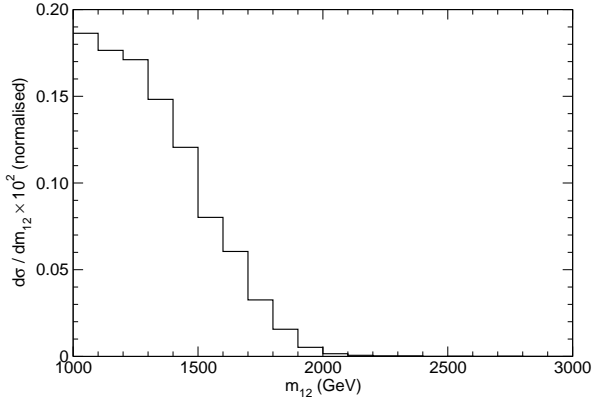}
\caption{Sample distribution for a quadriboson signal after event selection (see the text), normalised to unity. }
\label{fig:dist-quad}
\end{center}
\end{figure}

For quadriboson cascade decays as the one in figure~\ref{fig:diag2} the invariant mass distribution of the decay products $V_1$ and $V_2$ is also very wide at the partonic level, and less prone to shaping by kinematical cuts. We consider for simplicity that the masses of the two intermediate particles are equal, $M_{Y_1} = M_{Y_2} \equiv M_Y$, and that the extra particle $X$ is the same in both decays. The invariant mass distribution for $M_R = 2.25$ TeV, $M_Y = 650$ GeV, $m_1 = M_W$, $m_2 = M_Z$, $m_X = 100$ GeV is shown in figure~\ref{fig:dist-quadp}. Neglecting again width effects and working at first order in $m_1^2$, $m_2^2$ and $m_X^2$, the minimum and maximum values are
\begin{align}
& (m_{12}^2)^\text{min} = \left[ M_R^2 \, \frac{1 - \sqrt{1-4 \, M_Y^2/M_R^2}}{2} -M_Y^2 \right] \left( 1- \frac{2 m_X^2}{M_Y^2} \right) + m_1^2 + m_2^2 \,, \notag \\
& (m_{12}^2)^\text{max} = \left[ M_R^2 \, \frac{1+\sqrt{1-4 \, M_Y^2/M_R^2}}{2} -M_Y^2 \right] \left( 1- \frac{2 m_X^2}{M_Y^2} \right) + m_1^2 + m_2^2 \,.
\end{align}
As example of a quadriboson signal shape we take the distribution in figure~\ref{fig:dist-quad}, obtained with the high-purity $WZ$ event selection of the CMS Run 2 diboson search in the hadronic final state ~\cite{Sirunyan:2016cao}. As anticipated, the shaping is not very significant for quadriboson signals and for this event selection the only visible difference with the distribution at the partonic level is the cut of the distribution at $m_{12} = 1$ TeV. This example does not display any maximum at the range of invariant masses of interest. For other heavy resonance and/or intermediate resonance masses, the maximum could be within the invariant mass interval studied, but we expect that even in that case the effect would be similar because the width of the quadriboson bumps is really large, more than for triboson bumps. 

\section{Analysis framework}
\label{sec:3}

In many searches for narrow resonances decaying into two objects the background smoothly decreases with the invariant mass of the two objects (in our case of interest the diboson pair), but cannot be accurately predicted by Monte Carlo calculations. In these cases the background prediction in the signal region can be obtained by a fit of the observed distribution to a smoothly decreasing function --- with numerous checks in control and validation regions to see that the functional form proposed correctly describes data. 
For diboson resonance searches decaying into two fat jets $J$ the CMS Collaboration uses an empirical functional form~\cite{Khachatryan:2014hpa}
\begin{equation}
\frac{dN}{dm_{JJ}} = \frac{P_0 (1 - m_{JJ}/\sqrt s)^{P_1}}{(m_{JJ}/\sqrt s)^{P_2}} \,,
\label{ec:f3P}
\end{equation}
with $m_{JJ}$ the dijet invariant mass, $\sqrt s$ the centre of mass energy, and $N$ the number of events. The parameters $P_1$ and $P_2$ determine the shape, while $P_0$ is a normalisation factor.\footnote{In the latest analysis of Run 2 data~\cite{CMS:2016mwi} a simplified function without the $(1 - m_{JJ}/\sqrt s)^{P_1}$ factor in the numerator is used in some cases, and a generalised 5-parameter function obtained by replacing $P_2 \to P_2 + P_3 \log (m_{JJ}/\sqrt s) + P_4 \log^2 (m_{JJ}/\sqrt s)$ is required in some specific data sample.} The ATLAS Collaboration uses an equivalent functional form. The free parameters are fitted from data by maximising the binned likelihood function
\begin{equation}
L = \prod_{i} \frac{e^{-b_i} b_i^{n_i}}{n_i!} \,,
\label{ec:Lbkg}
\end{equation}
where $i$ runs over the different bins with numbers of events $n_i$, and $b_i$ is the predicted number of events (in this case background only) in each bin, obtained by integrating the function~(\ref{ec:f3P}) over the appropriate range.

Either when $b_i$ is predicted by Monte Carlo or in the above mentioned situation where it is obtained from a likelihood fit, the search for a peak on top of a smoothly falling background can be done by using a likelihood function,
\begin{equation}
L(\mu) = \prod_{i} \frac{e^{-(b_i+ \mu s_i)} (b_i + \mu s_i) ^{n_i}}{n_i!} \,,
\label{ec:L}
\end{equation}
with $s_i$ the predicted number of signal events in each bin, and $\mu$ a scale factor. For simplicity we omit additional terms in the likelihood that depend on nuisance parameters corresponding to systematic uncertainties, which are also fitted when maximising the likelihood. The probability density function (p.d.f.) of the $s_i$ used by the ATLAS and CMS Collaborations in diboson resonance searches is a single- or double-sided Crystal Ball function~\cite{CB,wikiCB}, which is a Gaussian where one or the two exponentially-decreasing tails are replaced by power-law tails to take into account the effects of radiation. When maximising the likelihood~(\ref{ec:L}), both $\mu$ and the free parameters in the background prediction are allowed to float. $P$-values are obtained by comparing the likelihood of a background-only distribution, given by eq.~(\ref{ec:f3P}), with a background plus signal one, where the parameters in the background distribution are also varied.
In order to compute expected and observed upper limits on a signal, the $\text{CL}_\text{s}$ method~\cite{Read:2002hq} is used, often with the asymptotic approximation of ref.~\cite{Cowan:2010js}.

\begin{figure}[htb]
\begin{center}
\includegraphics[height=5.25cm,clip=]{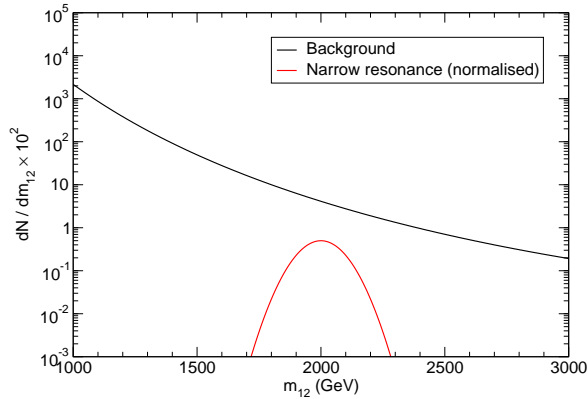}
\caption{Background distribution used as benchmark for our analysis, and p.d.f. distribution (normalised) of a 2 TeV narrow resonance.}
\label{fig:dist-CMS}
\end{center}
\end{figure}

We apply the above mentioned procedure to estimate the sensitivity to triboson and quadriboson bumps in data. First, we need a realistic assumption of a smoothly decreasing SM background. We take as example the distributions measured by CMS in the latest Run 2 analysis~\cite{Sirunyan:2016cao} with a luminosity of 12.9 fb$^{-1}$, specifically the data for the $WZ$ selection in high purity. As neither the fitted values for $P_0$, $P_1$ and $P_2$ nor the numbers of observed events in data are provided, a digitising tool~\cite{digitise} is used to estimate the numbers of events per bin (of variable width) and subsequently obtain the best-fit function, which is given by $\log P_0 = - 25.50$, $P_1 = -13.44$, $P_2 = 10.72$. This function, plotted in figure~\ref{fig:dist-CMS}, agrees very well with the best-fit function obtained in ref.~\cite{CMS:2016mwi}. The p.d.f. of the potential narrow resonance signals with centre $M$ (i.e. the resonance mass probed) has a standard deviation of $0.04 M$. The distributions shown by the CMS Collaboration seem quite close to a Gaussian shape, and the deviations from parabolae (in logarithmic scale) only occur at the tails where the value of the p.d.f. is two orders of magnitude smaller than the maximum. Therefore, and also because the additional parameters entering the Crystal Ball functions are not provided, in our analysis we take for simplicity Gaussian distributions of centre $M$ and standard deviation $0.04 M$. The probe function for $M = 2$ TeV is also shown in figure~\ref{fig:dist-CMS}.

%


\section{Sensitivity to multiboson resonances}
\label{sec:4}

In order to address the significance of the potential excesses, we inject signals corresponding to the three distributions in figures~\ref{fig:dist-trib} and~\ref{fig:dist-quad} over the assumed background in figure~\ref{fig:dist-CMS}. We perform a pseudo-experiment and compute the local $p$-value for different resonance masses, in intervals of 100 GeV, as well as the expected and observed 95\% upper limit on signal cross sections. We also calculate the $p$-value and upper limits for the Asimov dataset\footnote{Following ref.~\cite{Cowan:2010js} we denote the Asimov dataset as the one where the observed data correspond to the mean of the corresponding distributions, bin by bin.} for illustration, in order to find the deviations that one would generically expect, and also to show that the pseudo-experiments are not fine-tuned.

\subsection{Symmetric bump}
\begin{figure}[p]
\begin{center}
\begin{tabular}{cc}
\includegraphics[height=5.25cm,clip=]{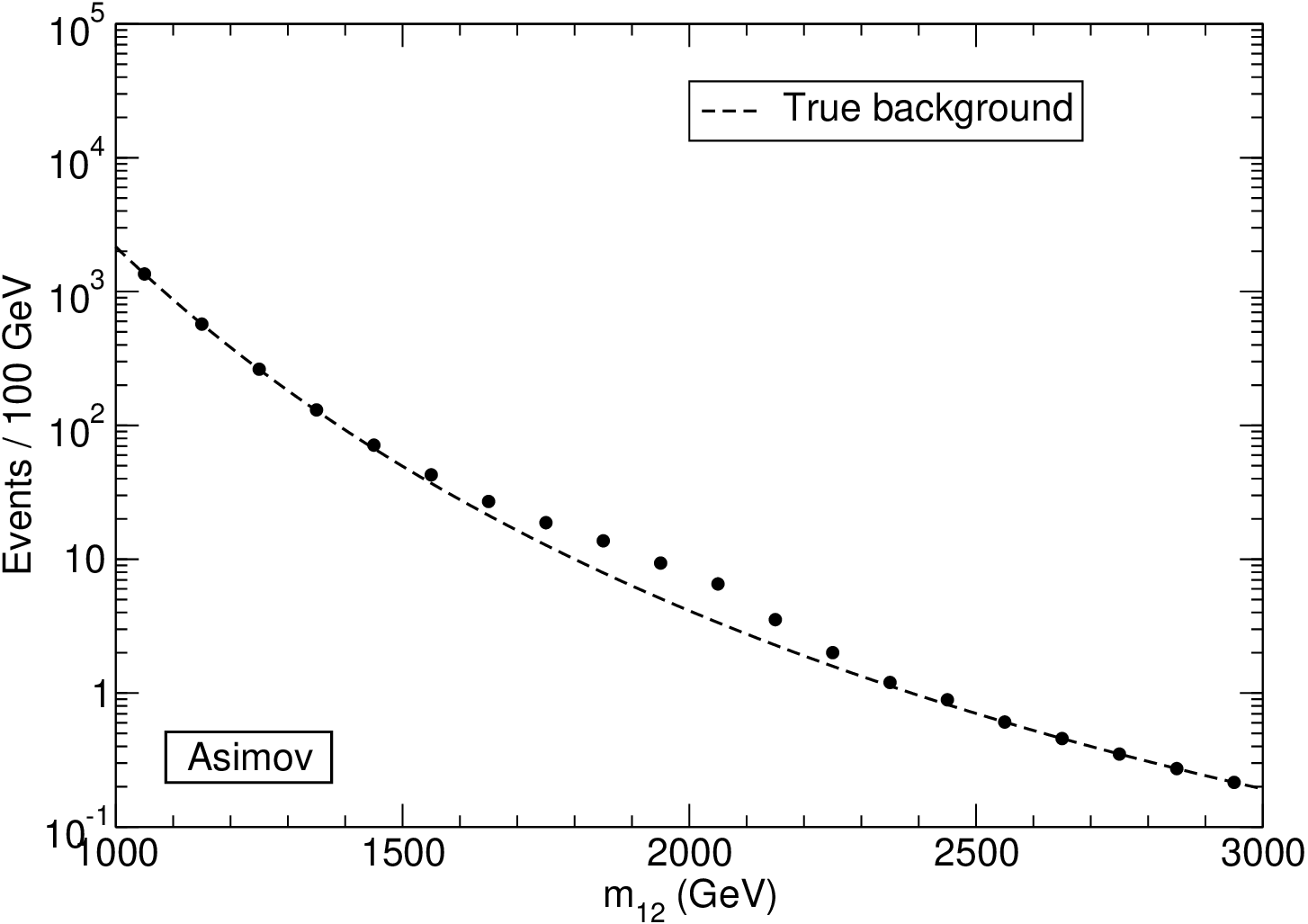} &
\includegraphics[height=5.25cm,clip=]{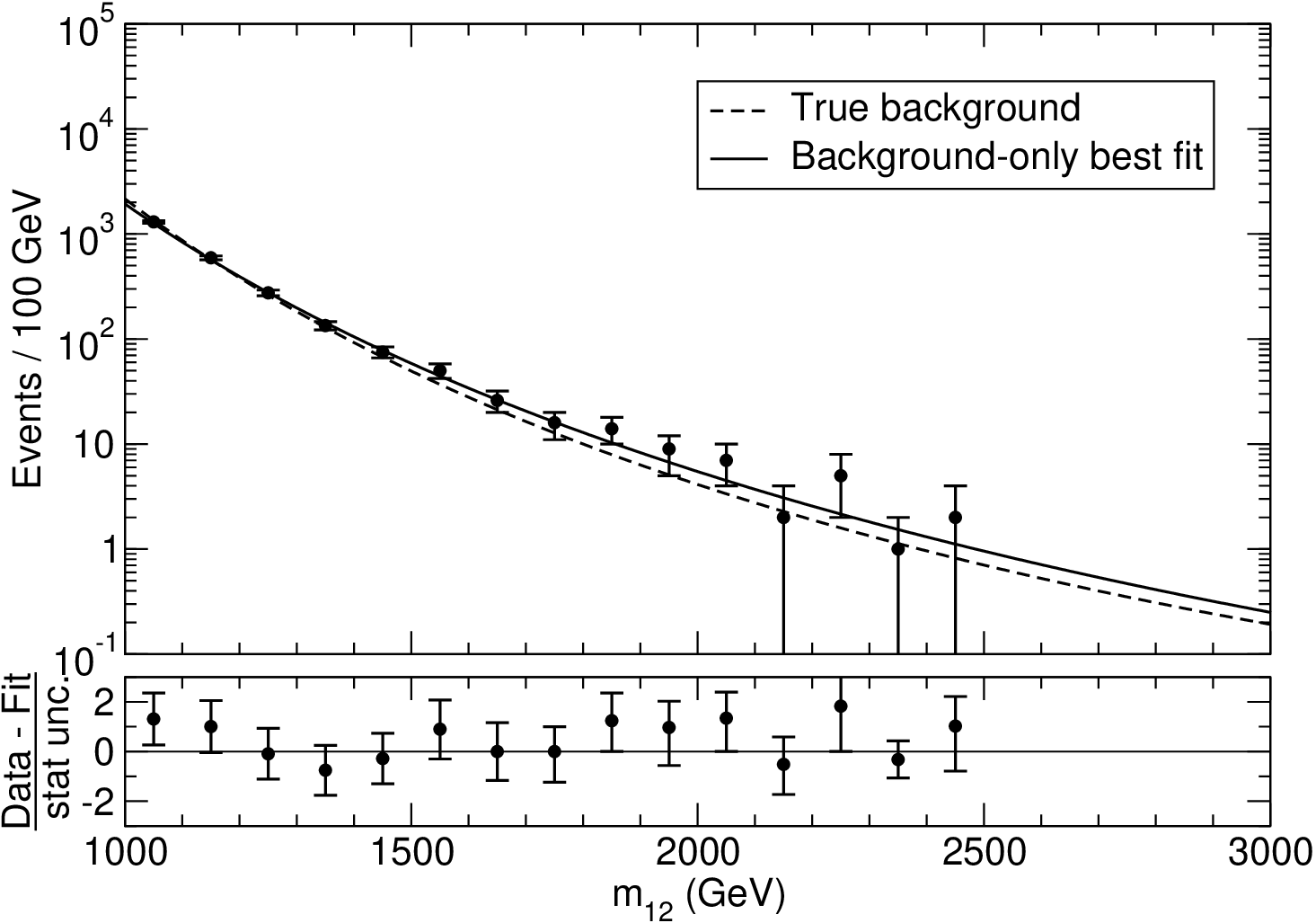} \\
\includegraphics[height=5.25cm,clip=]{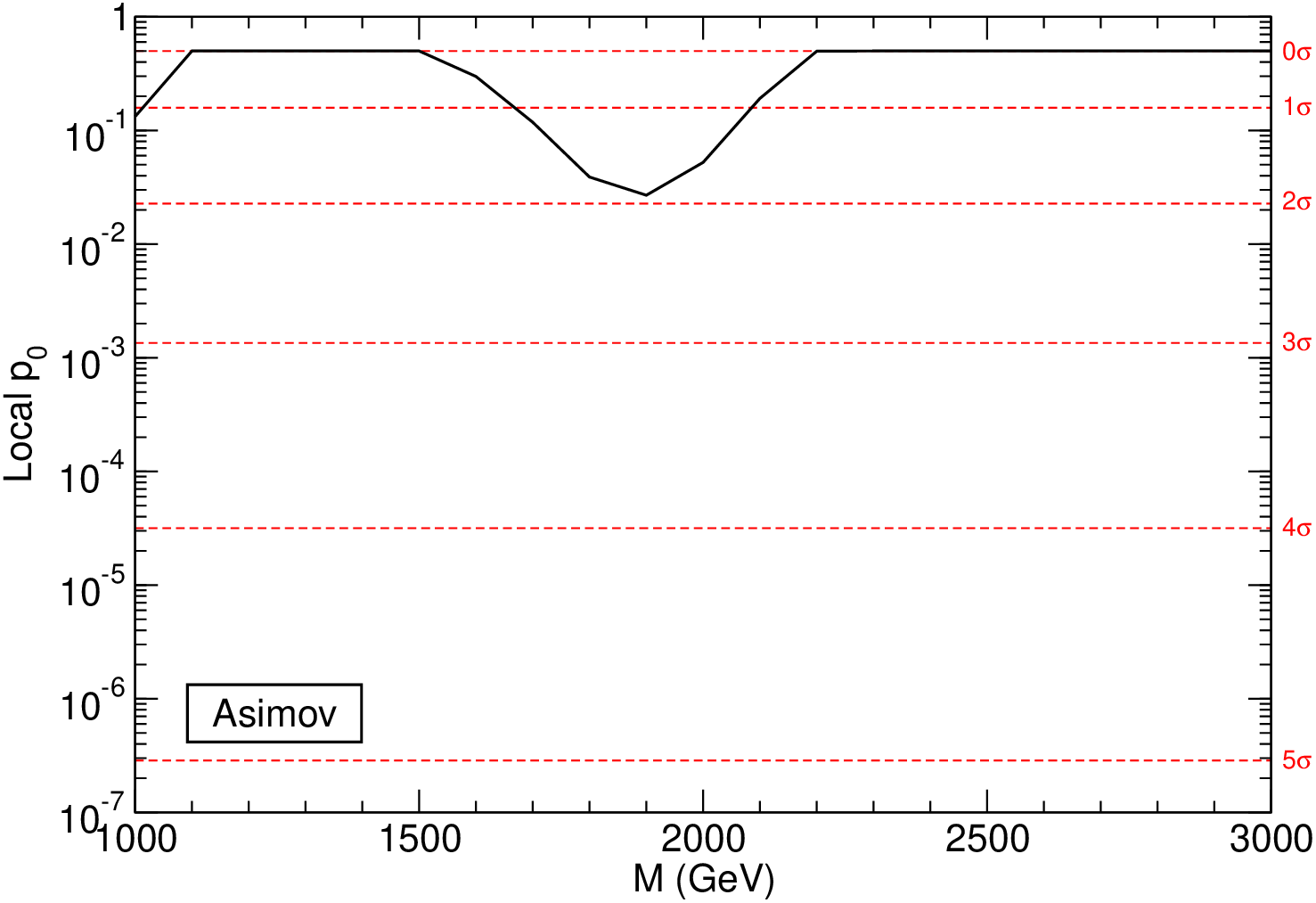} &
\includegraphics[height=5.25cm,clip=]{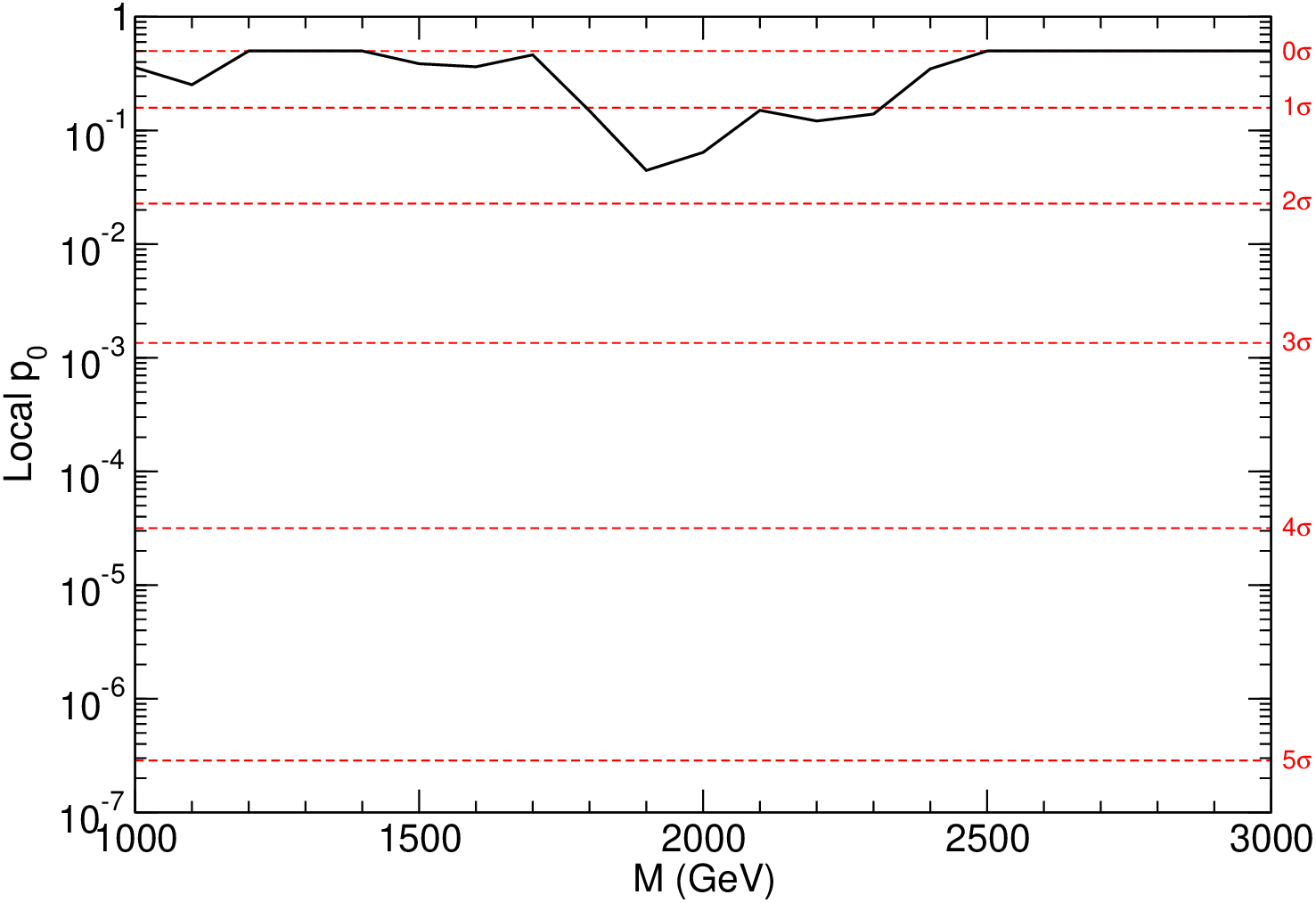} \\
\includegraphics[height=5.25cm,clip=]{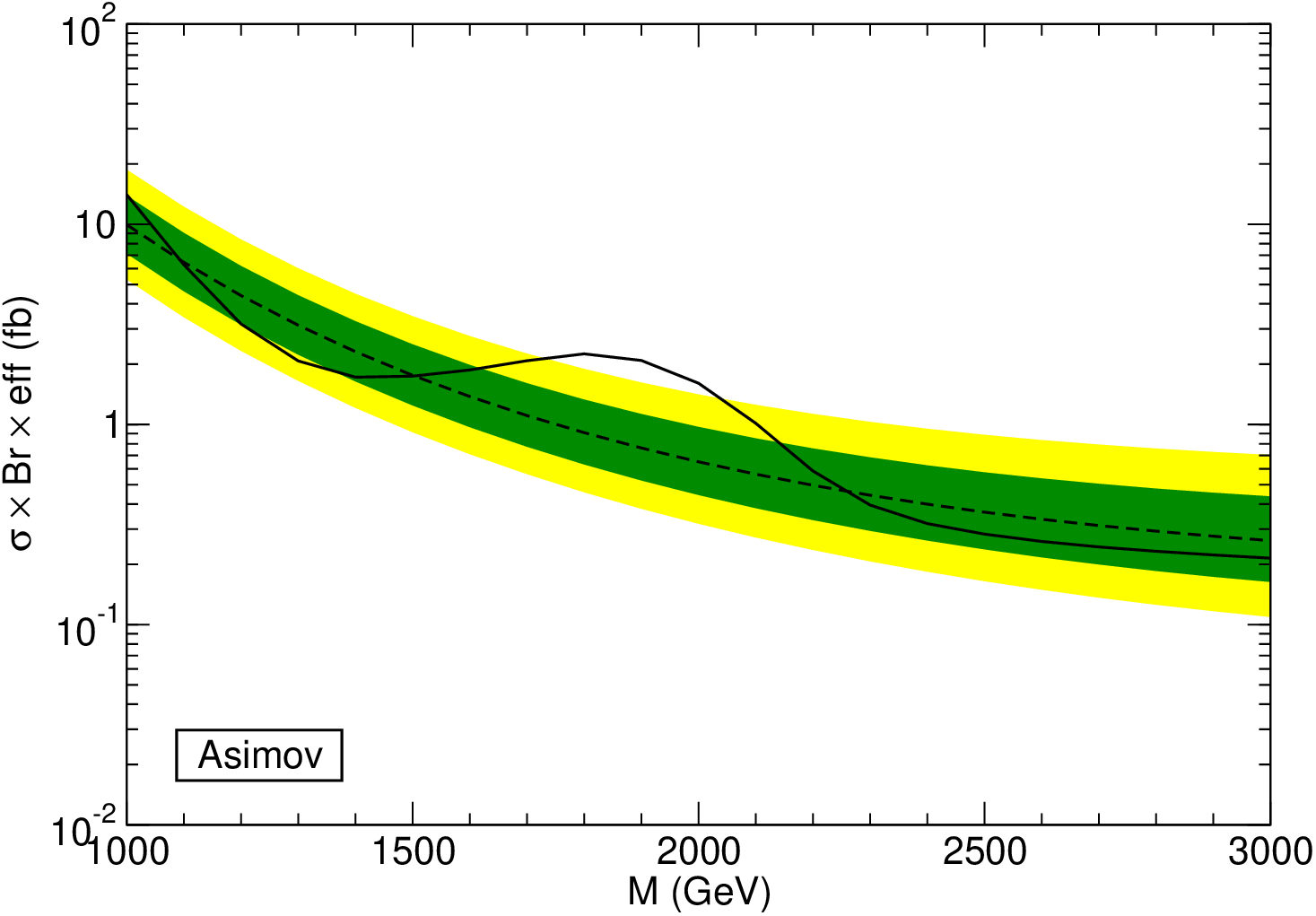} &
\includegraphics[height=5.25cm,clip=]{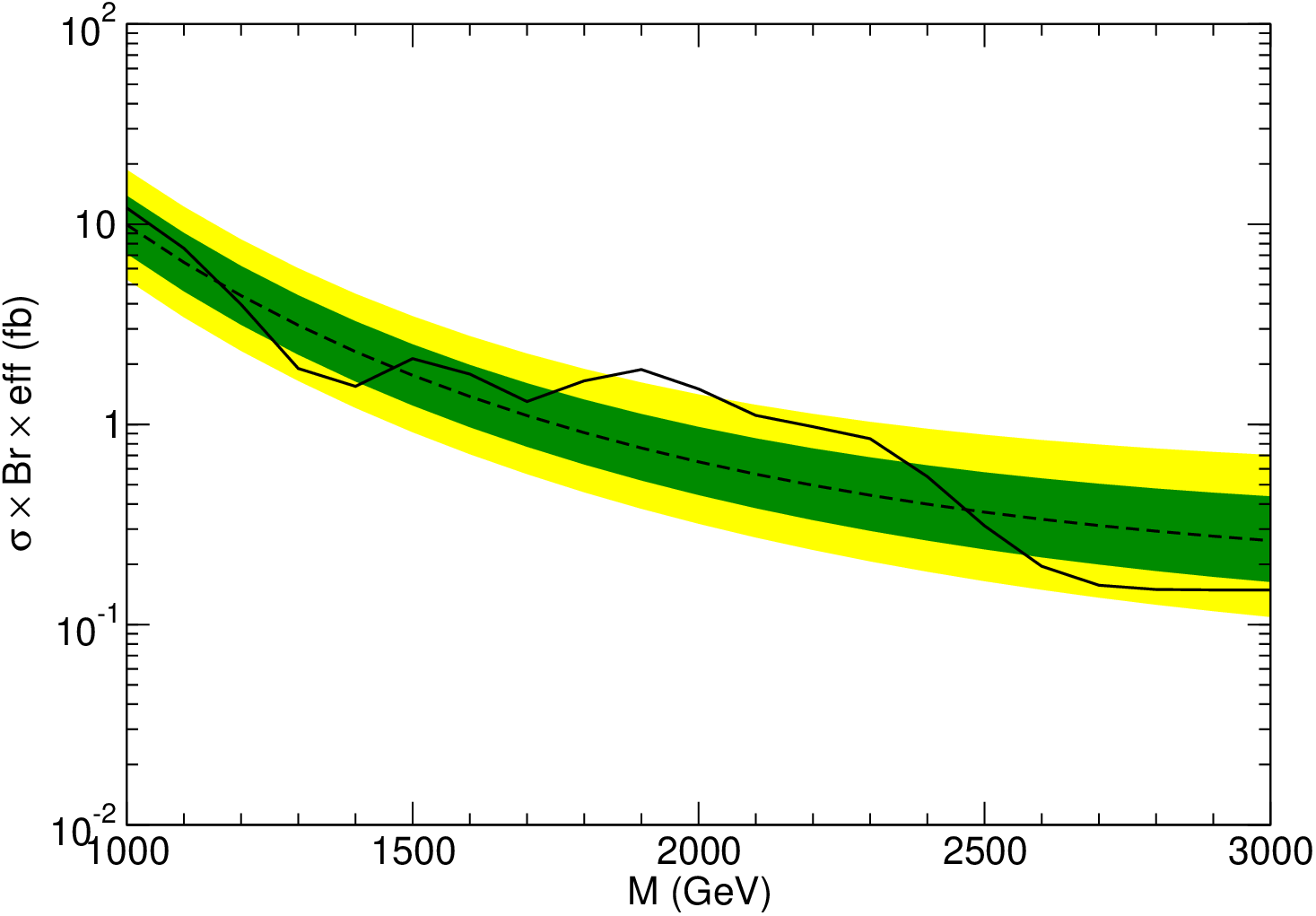} 
\end{tabular}
\caption{Top, left: Asimov data for the symmetric triboson bump in figure~\ref{fig:dist-trib} (left), with a size of 40 signal events. Top, right: data for a pseudo-experiment. Middle, left: Local $p$-values for the Asimov data. Middle, right: Local $p$-values for the pseudo-experiment. Bottom, left: Expected and observed limits for the Asimov data. Bottom, right: Expected and observed limits for the pseudo-experiment.}
\label{fig:bump1-data}
\end{center}
\end{figure}

For this distribution we add a signal with a mean of 40 events. The expected number of signal events in the selected invariant mass intervals $[1,3.5]$ TeV, $[1.5, 2]$ TeV and $[1.8, 2]$ TeV is given in table~\ref{tab:SB}.
The Asimov dataset is plotted in the top left panel of figure~\ref{fig:bump1-data}, together with the background.  A pseudo-experiment (i.e. a random sample) is given on the top right panel, with the background-only best fit as a solid line and the true background as a dashed line. We observe that the background-only best fit deviates from the background used as input, and absorbs part of the excess given by the injected signal. (This also happens, but to a lesser extent, in the presence of narrow resonances.) The best-fit parameters for the background-only hypothesis are given in table~\ref{tab:par} for the Asimov dataset. In this particular case, $P_2$ remains almost the same and the change in shape due to the injected signal is absorbed in $P_1$, shifting up the tail of the distribution.

\begin{table}[htb]
\begin{center}
\begin{tabular}{lccc}
& $1 - 3.5$ TeV & $1.5 - 2$ TeV & $1.8 - 2$ TeV \\
Background & 2562 & 85 & 13 \\
Symmetric bump & 40 & 32 & 10 \\
Asymmetric bump & 50 & 33 & 10 \\
Quadriboson tail & 100 & 31 & 2 \\
Narrow Gaussian & 30 & 30 & 7 \\
Wide Gaussian & 33 & 28 & 10
\end{tabular}
\caption{Number of mean background and signal events in selected invariant mass intervals.}
\label{tab:SB}
\end{center}
\end{table}

\begin{table}[htb]
\begin{center}
\begin{tabular}{lccc}
& $\log P_0$ & $P_1$ & $P_2$  \\
Background & -25.50 & -13.44 & 10.72 \\
Symmetric bump & -26.22 & -18.22 & 10.83 \\
Asymmetric bump & -25.95 & -17.61 & 10.74 \\
Quadriboson tail & -22.25 & -7.24 & 9.63 \\
Narrow Gaussian & -25.75 & -16.52 & 10.70 \\
Wide Gaussian & -26.13 & -17.69 & 10.81
\end{tabular}
\caption{Background-only best-fit parameters for the functional form in eq.~(\ref{ec:f3P}) for the input background and with various injected signals, for the Asimov datasets.}
\label{tab:par}
\end{center}
\end{table}

The presence of the bump cannot be seen by directly looking at the data points. The injected signal gives three consecutive bins around 2 TeV with a number of events $1\sigma$ larger than expected from the background-only best-fit function. In the example under study we know, by comparing with the Asimov dataset on the top left panel, that these mild excesses around 2 TeV are due to the injected signal. But when analysing real data one cannot determine if there is a real bump or merely a few statistical fluctuations.

The local $p$-values for the Asimov dataset and the pseudo-experiment are given in the middle row of figure~\ref{fig:bump1-data}. Although the injected number of signal events is quite large, namely 10 events over a background of 13 in the range $1.8 - 2$ TeV, the $p$-values are moderate and below the $2\sigma$ level. Notice that the smallest $p$-value is not found at the mean of the injected signal distribution, but is shifted to higher invariant masses. We also point out the small $\sim 1\sigma$ deviation in the first bin, where no signal events are present and the background is very large. This deviation is due to the above mentioned shift in the background-only best-fit function caused by the bump.

The bottom row of figure~\ref{fig:bump1-data} shows the expected and observed 95\% upper limits on signal cross section, for the Asimov dataset and the pseudo-experiment. (The expected limits and their uncertainty are computed using the background only.) The bottom left panel shows the generic expectation for a signal of this size, and the bottom right panel shows the result of the pseudo-experiment, which shows no evidence of a signal either. We remark that the presence of deficits in data, common even in the absence of a signal due to underfluctuations of the background, is slightly amplified by the upward shift in the background-only best fit caused by the wide bump.

\subsection{Asymmetric bump}

\begin{figure}[p]
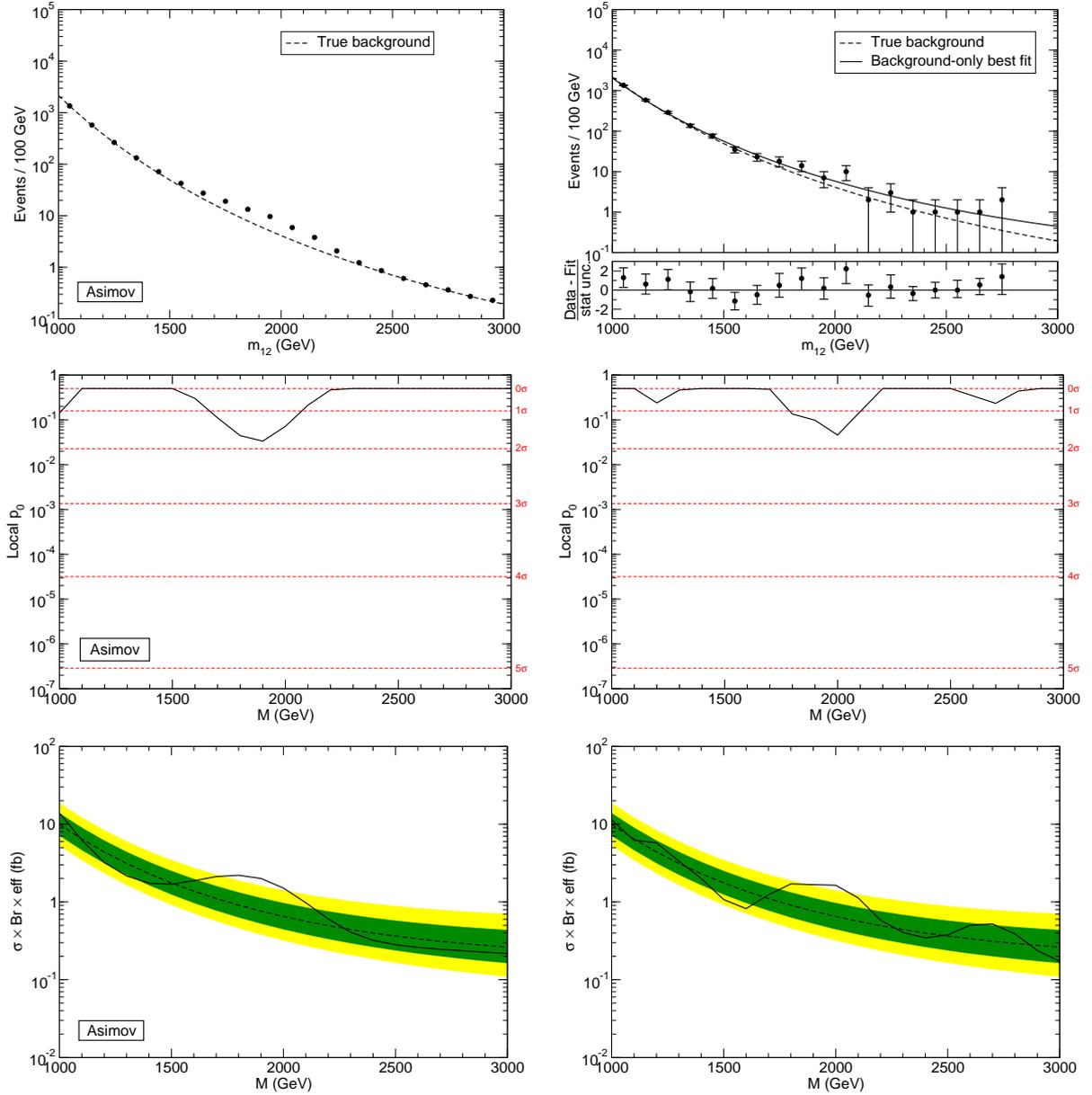

\begin{center}
\begin{tabular}{cc}
\includegraphics[height=5.25cm,clip=]{Figs/bump2-Asimovdata.eps} &
\includegraphics[height=5.25cm,clip=]{Figs/bump2-data.eps} \\
\includegraphics[height=5.25cm,clip=]{Figs/bump2-AsimovPval.eps} &
\includegraphics[height=5.25cm,clip=]{Figs/bump2-ObsPval.eps} \\
\includegraphics[height=5.25cm,clip=]{Figs/bump2-Asimovlimit.eps} &
\includegraphics[height=5.25cm,clip=]{Figs/bump2-Obslimit.eps} 
\end{tabular}
\caption{Top, left: Asimov data for the asymmetric triboson bump in figure~\ref{fig:dist-trib} (right), with a size of 50 signal events. Top, right: data for a pseudo-experiment. Middle, left: Local $p$-values for the Asimov data. Middle, right: Local $p$-values for the pseudo-experiment. Bottom, left: Expected and observed limits for the Asimov data. Bottom, right: Expected and observed limits for the pseudo-experiment.}
\label{fig:bump2-data}
\end{center}
\end{figure}

Here we add a larger signal, with a mean of 50 events. The expected number of signal and events in selected invariant mass intervals is given in table~\ref{tab:SB}, and coincides with the symmetric bump for invariant masses in the ranges $[1.5,2]$ TeV and $[1.8,2]$ TeV. The Asimov dataset is shown in the top left panel of figure~\ref{fig:bump2-data}, and the best-fit parameters in table~\ref{tab:par}. The top right panel displays the result of a pseudo-experiment. One can see that the shift in the background-only best-fit function, that is, the difference between the dashed and solid lines, is larger than for the symmetric bump. This larger shift is caused by the low-mass tail of the asymmetric bump. Again, the presence of the bump cannot be seen by directly looking at the data points. By comparing with the Asimov dataset we see that the mild excess events around 2 TeV are due to the bump, but in real data these would merely appear as statistical fluctuations.

The local $p$-values are given in the middle row, and the expected and observed 95\% upper limits on signal cross section in the bottom row.
Altogether, the results for the asymmetric bump are quite similar to the ones for the symmetric bump but with smaller deviations for the number of signal events assumed, which coincide in the relevant invariant mass ranges $[1.5,2]$ TeV and $[1.8,2]$ TeV. This small difference is caused by the slightly larger shift of the background-only best-fit function. We notice again that the minimum $p$-value is not found at the maximum of the injected signal, but is moved to higher masses. The deficits in data are also present, and amplified by the shift in the background-only best fit, which also produces a small $\sim 1\sigma$ deviation in the first bin.

\subsection{Quadriboson tail}

\begin{figure}[p]
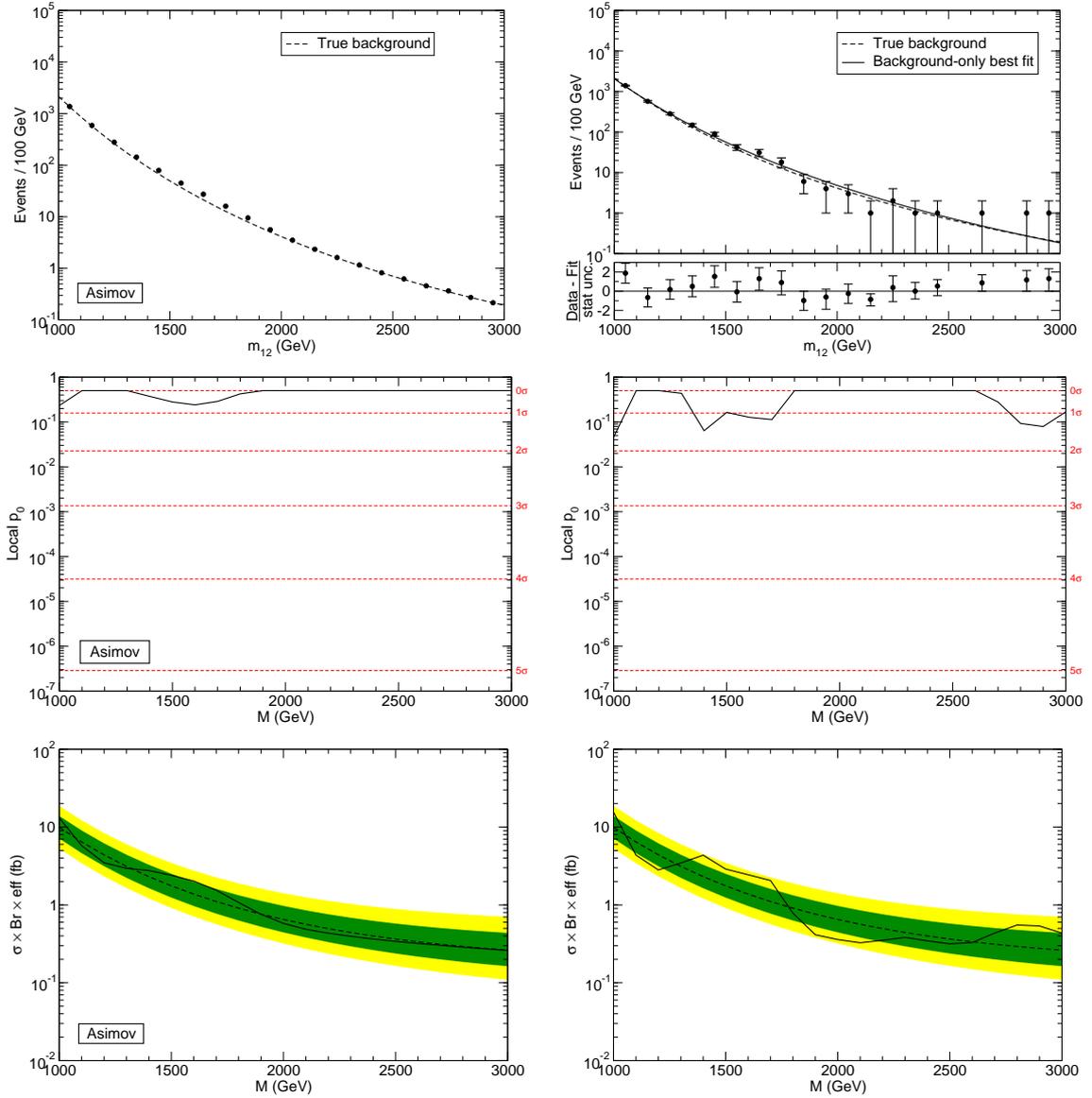

\begin{center}
\begin{tabular}{cc}
\includegraphics[height=5cm,clip=]{Figs/bump3-Asimovdata.eps} &
\includegraphics[height=5cm,clip=]{Figs/bump3-data.eps} \\
\includegraphics[height=5cm,clip=]{Figs/bump3-AsimovPval.eps} &
\includegraphics[height=5cm,clip=]{Figs/bump3-ObsPval.eps} \\
\includegraphics[height=5cm,clip=]{Figs/bump3-Asimovlimit.eps} &
\includegraphics[height=5cm,clip=]{Figs/bump3-Obslimit.eps} 
\end{tabular}
\caption{Top, left: Asimov data for the quadriboson tail in figure~\ref{fig:dist-quad}, with a size of 100 signal events. Top, right: data for a pseudo-experiment. Middle, left: Local $p$-values for the Asimov data. Middle, right: Local $p$-values for the pseudo-experiment. Bottom, left: Expected and observed limits for the Asimov data. Bottom, right: Expected and observed limits for the pseudo-experiment.}
\label{fig:bump3-data}
\end{center}
\end{figure}

We inject a signal with a mean of 100 events, twice larger than in the previous cases. The number of expected events in the relevant invariant mass intervals, where the background is small, is given in table~\ref{tab:SB}. The Asimov dataset and the pseudo-experiment are shown in the top left and top right panels of figure~\ref{fig:bump3-data}, respectively. It is apparent that the differences in shape with respect to the assumed background are minimal, and the assumed background distribution fits very well the data points even if both $P_1$ and $P_2$ deviate significantly from the initial values without injected signal, see table~\ref{tab:par}. This is also reflected in the local $p$-values in the middle row. 
For the Asimov dataset, the $p$-values are always close to $1/2$, and for the pseudo-experiment the deviations are merely caused by fluctuations in pseudo-data. The same argument holds for the observed limits on signal cross section in the bottom panels: the observed limit is always within the $2\sigma$ band and the variations are caused by statistical fluctuations. This kind of signal is practically invisible in these searches.

\subsection{Comparison with Gaussian shape resonances}

\begin{figure}[t]
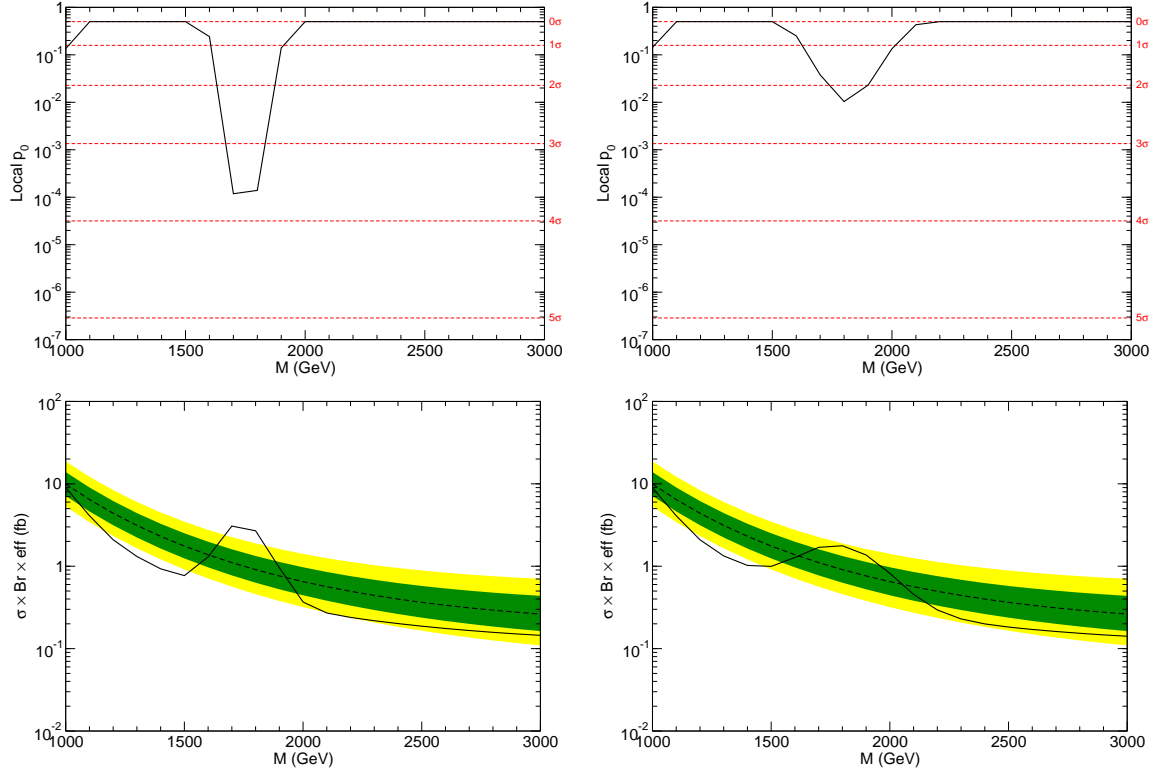

\begin{center}
\begin{tabular}{cc}
\includegraphics[height=5cm,clip=]{Figs/Gauss1-AsimovPval.eps} &
\includegraphics[height=5cm,clip=]{Figs/Gauss2-AsimovPval.eps} \\
\includegraphics[height=5cm,clip=]{Figs/Gauss1-Asimovlimit.eps} &
\includegraphics[height=5cm,clip=]{Figs/Gauss2-Asimovlimit.eps} 
\end{tabular}
\caption{Top: Local $p$-values for the Asimov data, for the narrow (left) and wide (right) Gaussian shapes. Bottom: Expected and observed limits for the Asimov data, for the narrow (left) and wide (right) Gaussian shapes.}
\label{fig:gauss-data}
\end{center}
\end{figure}

In order to put in context the (in)sensitivity of these searches to wide bumps, we compare with the deviations one would expect for a narrow and a wide resonance with Gaussian shape, with $M = 1.75$ TeV and $\Gamma = 0.04 M$ and $\Gamma = 0.1 M$, respectively. For the narrow resonance we assume a mean of 30 events, and for the wide resonance a mean of 33 events. The expected numbers of events in the invariant mass ranges of interest are collected in table~\ref{tab:SB}. Figure~\ref{fig:gauss-data} shows the local $p$-values and observed limits for the Asimov dataset in both cases. The best-fit parameters are given in table~\ref{tab:par}.

Clearly, a narrow Gaussian is very visible in data, with a minimum $p$-value that reaches $3.7\sigma$ and an observed upper limit that significantly deviates from the $2\sigma$ band. The wide Gaussian resonance is similar to the symmetric triboson bump, but it is more visible, with a smaller $p$-value ($2.3\sigma$ versus $1.9\sigma$ for the Asimov datasets). Also, with a wide Gaussian resonance the deficits in data, clearly seen by points where the observed limit is much smaller than the expected one, are likely to be more pronounced than for a triboson bump.

\section{Summary and discussion}
\label{sec:5}

In this paper we have addressed the visibility of new physics signals that give a wide excess over a smoothly falling background that cannot be accurately predicted. We have taken as example the signals that triboson and quadriboson resonances would give in dibsoson searches in hadronic final states. But, clearly, the scope and applications of our study are much broader.

We have selected three shapes for signal distributions as benchmarks for our study: a symmetric bump and an asymmetric bump, both resulting from triboson resonances, and a quadriboson tail. Our results can be summarised as follows.
\begin{enumerate}
\item The symmetric bump is similar to a wide resonance of Gaussian shape, but less visible, if we compare signals of the same size. For the symmetric bump both the $p$-value and the 95\% upper limit on signal cross section are seen to deviate less from the background-only expectation.
\item The asymmetric bump is even less visible due to its low-mass tail. The quadriboson tail is practically invisible in these searches, even for relatively large numbers of signal events.
\item As it is obvious, the presence of a wide bump on a smootly falling background changes the ``background-only best fit'' with respect to the ``true'' background. Therefore, mild deficits in data are expected at both sides of the bump. These deficits are not detected by the local $p$-value, which by construction gives low values only for positive deviations in data. In case that the data also underfluctuates at the sides of the bump the dips may be significant, and can be seen in any case by comparing the expected and observed limits.
\item The shift in the background-only best fit mentioned above can also give rise to a $\sim 1\sigma$ excess in the leftmost bins, as it happens in the studied examples. This is also easy to understand, since the convexity of the best-fit curve may decrease by the bump. 
\end{enumerate}

One may also wonder whether, independently of the statistical estimators of the data compatibility with the background-only hypothesis, one could detect by eye the presence of wide bumps. In order to do so, sufficient statistics are required. In the pseudo-experiments shown it is hard, not to say impossible, to determine only by eye if the pseudo-data exhibit a bump or, on the contrary, we merely have a few upper statistical fluctuations at the $1\sigma$ level in adjacent bins. In our study we have taken as background a function that approximately reproduces the high-purity $WZ$ sample in the CMS diboson search with 12.9 fb$^{-1}$, and we have seen that triboson signals of 50 events (after branching ratios and efficiency factors) would remain unseen. Quadriboson signals of a much larger size would be invisible as well. One can then expect that, in order to explore the 2 TeV region in detail, one would need five times more statistics. 

Our results highlight a weakness of searches in which the background cannot be accurately predicted by Monte Carlo calculations. Even if some of the searches are performed for wide resonances too, signals like the triboson bumps and quadriboson tails cannot be accomodated by resonances with a Gaussian shape. Therefore, the sensitivity to these new physics processes is small. We note that the signal distributions in this paper have been obtained by a realistic simulation~\cite{Aguilar-Saavedra:2016xuc}, and correspond to the generic profiles that one sees for these new physics signals. 

Reversing our arguments, the low sensitivity to this type of signals implies that they could exist in current LHC data, and yet be mistaken by small statistical fluctuations of a few adjacent bins, yielding a local $p$-value at the $2\sigma$ level, not statistically significant. Let us stress again that new heavy resonances at the reach of the LHC can undergo cascade decays and give multiboson signals while diboson signals are absent, as pointed out in ref.~\cite{Aguilar-Saavedra:2015iew}. Diboson resonance searches in the semileptonic decay modes have low sensitivities, because of the presence of the extra particles~\cite{Aguilar-Saavedra:2015rna,Aguilar-Saavedra:2016xuc}. Moreover, as shown in this work, in the absence of a significant signal shaping the hadronic decay modes have little sensitivity too. The low sensitivity of current analyses, with the absence of any convincing hint of a more ``conventional'' form of new physics beyond the SM, should motivate dedicated searches for triboson and quadriboson resonances in the LHC experiments.

\section*{Acknowledgements}
I thank J.C. Collins and S. Lombardo for previous collaboration, and J.C. Collins also for discussions on statistics. This work has been supported by MINECO Projects  FPA 2016-78220-C3-1-P and FPA 2013-47836-C3-2-P (including ERDF), and by Junta de Andaluc\'{\i}a Project FQM-101.

\end{document}